\documentclass[fleqn,12pt,twoside]{article}
\usepackage{espcrc1,epsfig,graphicx}

\usepackage{graphicx}
\usepackage[figuresright]{rotating}

\def\beq{\begin{equation}}
\def\eeq{\end{equation}}

\def\epm#1#2{\hbox{${\lower1pt\hbox{$\scriptstyle +#1$}}
\atop {\raise1pt\hbox{$\scriptstyle -#2$}}$}}

\def\gsim{\mathrel{\rlap{\lower4pt\hbox{\hskip1pt$\sim$}}
    \raise1pt\hbox{$>$}}}         

\def\frac#1#2{{{#1}\over {#2}}}
\def\half{\hbox{${1\over 2}$}}

\catcode`@=11 
\def\slash#1{\mathord{\mathpalette\c@ncel#1}}
 \def\c@ncel#1#2{\ooalign{$\hfil#1\mkern1mu/\hfil$\crcr$#1#2$}}
\def\lsim{\mathrel{\mathpalette\@versim<}}
\def\gsim{\mathrel{\mathpalette\@versim>}}
 \def\@versim#1#2{\lower0.2ex\vbox{\baselineskip\z@skip\lineskip\z@skip
       \lineskiplimit\z@\ialign{$\m@th#1\hfil##$\crcr#2\crcr\sim\crcr}}}
\catcode`@=12 

\def\be{\begin{equation}}
\def\ee{\end{equation}}
\def\bea{\begin{eqnarray}}
\def\eea{\end{eqnarray}}

\def\epm#1#2{\hbox{${\lower1pt\hbox{$\scriptstyle +#1$}}
\atop {\raise1pt\hbox{$\scriptstyle -#2$}}$}}

\newcommand{\AmS}{{\protect\the\textfont2
  A\kern-.1667em\lower.5ex\hbox{M}\kern-.125emS}}

\title{Structure Functions and Parton Distributions}

\author{Stefano Forte \address[SF]{Dipartimento di Fisica, Universit\`a di Milano and\\ 
INFN, Sezione di Milano, Via Celoria 16, I-20133 Milano, Italy}}       
\begin{document}
\begin{flushright}
{\tt hep-ph/0502073}
\end{flushright}
\begin{center}

{\Large\bf Structure Functions and Parton Distributions}\\
\bigskip
\bigskip
{\large\bf  Stefano Forte}\\

\bigskip
{ \it Dipartimento di Fisica, Universit\`a di Milano and\\ 
INFN, Sezione di Milano, Via Celoria 16, I-20133 Milano, Italy}
\bigskip
\bigskip
\bigskip
\vskip3cm

{\bf Abstract}\\
\end{center}
\bigskip
\noindent
I review recent progress in the determination of the parton structure
of the nucleon, in particular from  deep-inelastic
structure functions. I explain how the needs of current and future
precision phenomenology, specifically at the LHC, have turned the
determination of parton distributions into a quantitative problem. I
describe the results and difficulties of current approaches and
ideas to go beyond them.\hfill\\
\vskip3cm
\begin{center}
 Invited plenary talk at\\
{\bf BARYONS04}\\
Paris, October 2004\\
{\it to be published in the proceedings}
\end{center}
\vfill
February 2005\hfill IFUM-822/FT
\eject

\maketitle

\begin{abstract}
I review recent progress in the determination of the parton structure
of the nucleon, in particular from  deep-inelastic
structure functions. I explain how the needs of current and future
precision phenomenology, specifically at the LHC, have turned the
determination of parton distributions into a quantitative problem. I
describe the results and difficulties of current approaches and
ideas to go beyond them.
\end{abstract}

\section{From HERA to the LHC}
Knowledge of the parton structure of the nucleon has undergone a
revolution during the last decade, driven by present and
future experimental data. On the one hand, current experiments,
especially at HERA~\cite{herarev} 
but also from neutrino beams at Fermilab, have provided us with
an unprecedented amount of experimental information, mostly from the
measurement of  deep-inelastic structure functions. On the other hand,
LHC, now behind the corner, will require, essentially for the first
time, a precision approach to the structure of the nucleon in the
context of searches for new physics~\cite{lhcrev}. 
This has stimulated a considerable amount of theoretical and
phenomenological work, with the aim of turning the physics of parton
distributions into a quantitative science. 

\section{Determining PDFS}

\begin{figure}[ht]
\vspace{-1.cm}
\includegraphics[width=.52\linewidth,clip]{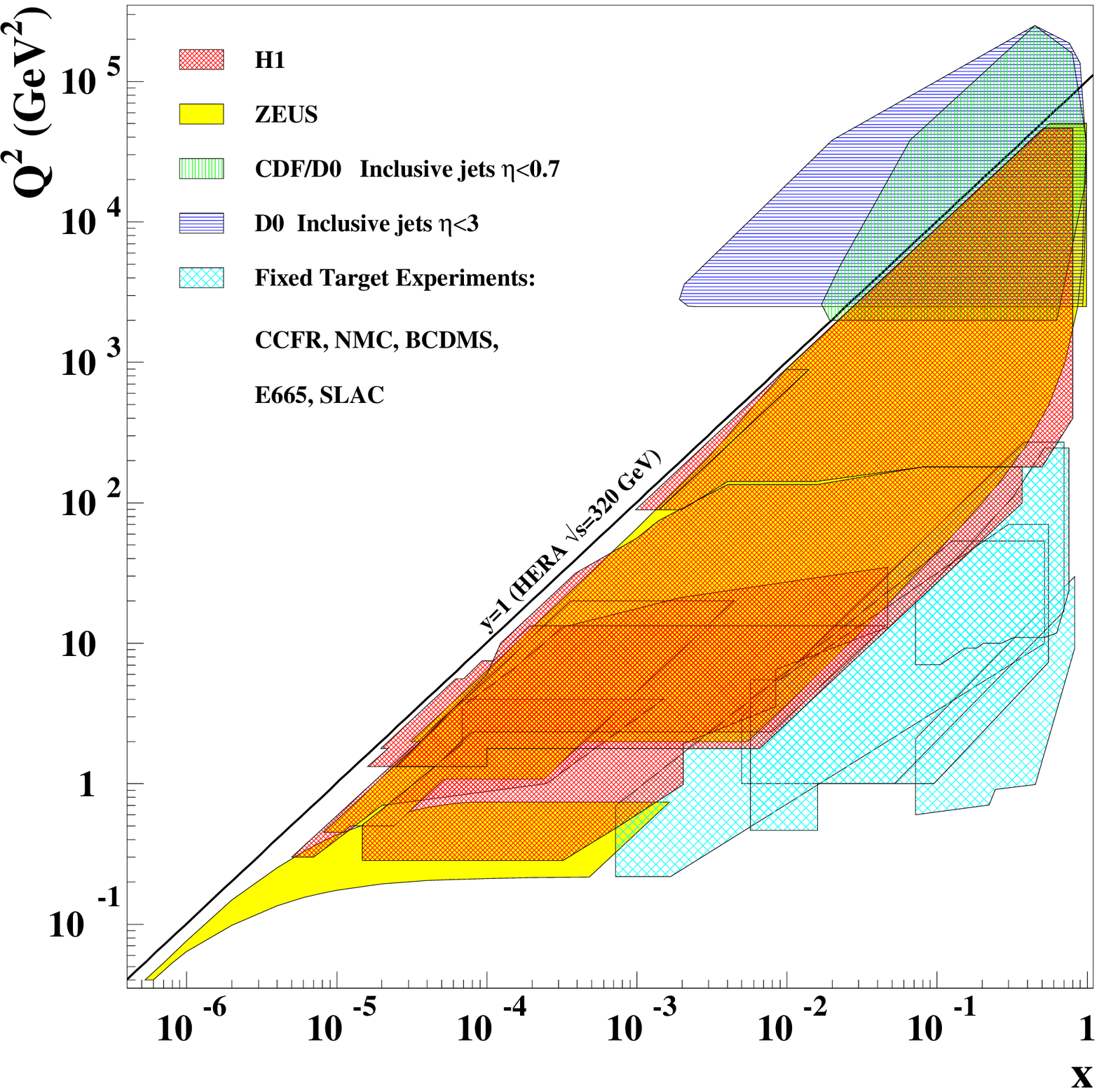}
\includegraphics[width=.47\linewidth,clip]{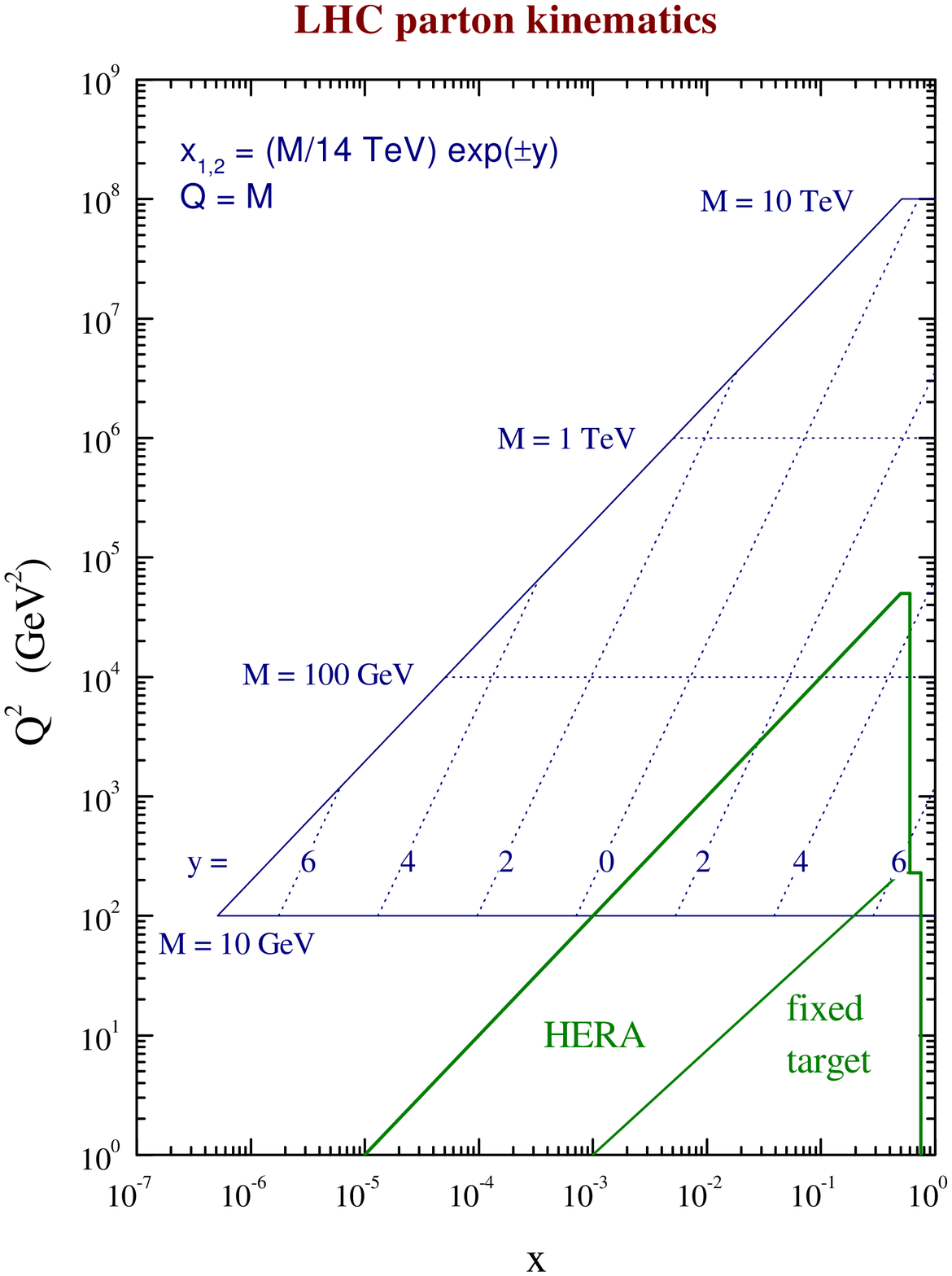}
\vskip-1.cm
\caption{Kinematic coverage of
  current data (left, from Ref.~\cite{herarev}) and the LHC (right,
  from Ref.~\cite{lhcrev}).} 
\label{kinreg}
\vspace{-.5cm}
\end{figure}
The parton structure of the nucleon can be determined thanks to
factorization: a physical cross section is expressed as the
convolution of perturbatively computable parton cross sections, times
parton densities (pdfs). We can then use one process to measure
pdfs, which are then used to compute a different process.
In the prototypical case of 
inclusive deep-inelastic scattering (DIS) the
cross section, up to  
corrections suppressed by powers of $m_p^2/Q^2$, is given by
\bea
\label{disxsec}
&&
\frac{d^2\sigma^{\lambda_p\lambda_\ell}(x,y,Q^2)}{dx dy}
=
\eta
\frac{Q^2}{ xy}\Bigg\{
\left[-\lambda_\ell\, y \left(1-\frac{y}{2}\right) x { F_3(x,Q^2)}
      \right.\nonumber\\ &&\quad\left.
+(1-y) { F_2(x,Q^2)} + y^2 x {
F_1(x,Q^2)}\right]
 -2\lambda_p
  \left[
     -\lambda_\ell\, y (2-y)  x { g_1(x,Q^2)}
\right.\nonumber\\ &&\qquad\left. -(1-y) {
g_4(x,Q^2)}- y^2 x { g_5(x,Q^2)}
  \right]
\Bigg\},
\eea
where $\lambda$ are the lepton and proton helicities (assuming
longitudinal proton polarization),  the
kinematic variables are $y=\frac{p\cdot q}{p\cdot
k}$ (lepton fractional energy loss), $x= \frac{Q^2}{2 p\cdot
q}$ (Bjorken $x$), and $\eta$ depends on the gauge bosons which
mediates the scattering process:
\beq
\eta_\gamma=\frac{4\pi\alpha^2}{Q^2};\quad
\eta_W=G^2_F\frac{Q^2}{  2\pi (1+Q^2/m_W^2)^2};\quad
\eta_Z=G^2_F  [\half(g_V-\lambda_\ell
g_A)]^2
\frac{Q^2}{  2\pi (1+Q^2/m_Z^2)^2}.
\label{etafac}
\eeq
More contributions are due  to interference of different exchange
processes.

The factorization theorem expresses the structure functions which
parametrize the cross section as a convolution of a perturbative
partonic cross section (coefficient function) and a pdf.
For example
\bea 
\label{f2}F_2^{\rm \gamma}(x,Q^2)&&=x \sum_{\hbox{\rm\scriptsize flav.} \>i}\Bigg\{ { e^2_i}{ \left(q_i(x,Q^2)+\bar
q_i(x,Q^2)\right)}\nonumber\\&&\qquad+{ \alpha_s(Q^2)} \left[{
    C_i[x,\alpha_s(Q^2)]}\otimes {\left(q_i(x,Q^2)+
\bar
q_i(x,Q^2)\right)}\right.\\
&&\qquad\qquad\left.
+{ C_g[x,\alpha_s(Q^2)]} \otimes {g}\right]\Bigg\},\nonumber
\eea
where $f(f)\otimes g(x)\equiv \int_x^1\frac{dy}{y} f(x/y) g(y)$ and
$C_1=1+ O(\alpha_s)$, $C_g=O(\alpha_s)$ are respectively the $i$-th
quark flavour and gluon coefficient functions, i.e. the perturbative
cross-sections
for the gauge boson-parton scattering process.
The structure function $F_1$ depends on the same combination of quarks
as $F_2$, but with a different gluon content:
\bea
&&F_L^{\rm \gamma}(x,Q^2)\equiv F_2^{\rm \gamma}(x,Q^2)- 2x F_1^{\rm
  \gamma}(x,Q^2)\nonumber\\&&\qquad=\sum_{\hbox{\rm\scriptsize flav.} \>i}
{ \alpha_s(Q^2)} \left[{
    C_i^L[x,\alpha_s(Q^2)]}\otimes {\left(q_i(x,Q^2)+
\bar
q_i(x,Q^2)\right)}+{ C_g^L[x,\alpha_s(Q^2)]} \otimes {g}\right].
\label{fl}
\eea
Other structure
functions are sensitive to different combinations of parton flavours:
the $g_i$ structure functions are spin--odd and contribute to the
polarized cross section; 
the structure functions 
$F_3$, $g_4$ and $g_5$ are parity--violating, and contribute to
weak current scattering. For unpolarized $\gamma^*$ DIS only $F_1$ and
$F_2$ contribute. 

\begin{figure}[ht]
\vspace{-1.5cm}
\includegraphics[width=.55\linewidth,clip]{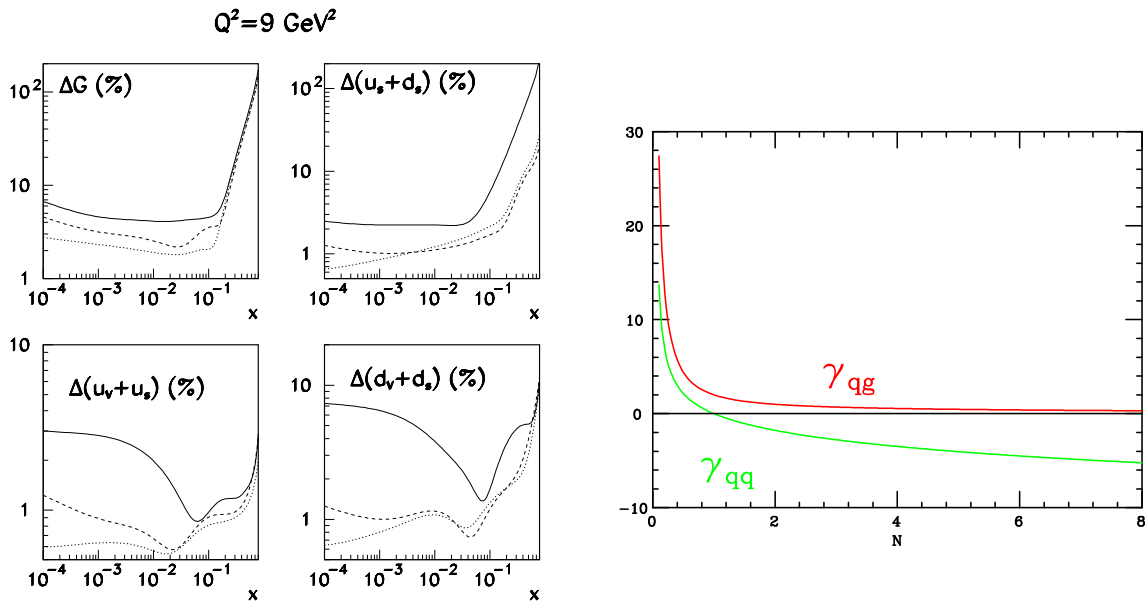}
\includegraphics[width=.44\linewidth,clip]{gs.ps}
\vskip-.8cm
\caption{Left: impact of Drell-Yan data: top (dots) DIS only;
middle (dashes) DIS+E866 Drell-Yan data; bottom (solid) DIS+ E866 and
projected LHC Drell-Yan data (from Ref.~\cite{aleprel})
Right: gluon anomalous dimensions in $N$--space. }
\label{gluplot}
\vspace{-.5cm}
\end{figure}

Various physical processes are given by different combinations of the same pdfs
convoluted with the appropriate partonic cross sections and kinematics. In
particular, for a process at a hadron collider (such as Higgs production at the
LHC)
where the hard scale is
the mass $M^2$ of the final state  
$Q^2=M^2$ and $x_i=\sqrt{\frac{M^2}{s}} \exp\pm y$, where
$s$ is the center-of-mass energy of the hadronic collision, $\pm y$ the
parton rapidities and $i$ refers to each of the incoming hadrons.
The kinematic regions for HERA and LHC are compared in fig.~1, along
with 
the current experimental coverage of the $(x,Q^2)$ plane from
unpolarized DIS data. In order to obtain
predictions for LHC processes we must solve three problems: 1) disentangle
the contribution of individual partons to the observable used to
determine them;  2) evolve
them up to the relevant scale and convolute them with the appropriate
perturbative cross section; 3) determine the error on them. 

\subsection{Phenomenology: disentangling PDFS}

Whereas deep-inelastic scattering mediated by $\gamma^*$ exchange
provides the bulk of the data shown in fig.~1 (most of the HERA data
and older fixed target data), they only measure the cross--section
eq.~(\ref{disxsec}), i.e., in turn, the
combination of
pdfs of eqs.~(\ref{f2}-\ref{fl}). This means that: 1) only the C--even combination
$q+\bar q$ is accessible; 2) flavour separation can be done only for $u$
and $d$ quark using proton and deuteron targets (and then only in
fixed target experiment, since a HERA upgrade with nuclear beams has
not  been approved); 3) the gluon contribution can only be determined
through scaling violations:
\beq
{ {d\over dt} F_2^s (N,Q^2)}=
{\alpha_s(Q^2)\over
2\pi} \left[{\gamma_{qq}(N)} { F_2^{s}}+2  \, 
n_f{\gamma_{qg}(N)g(N,Q^2)}\right]+O(\alpha_s^2),
\label{f2ev}
\eeq
where $N$ is related to $x$ by  Mellin transform 
according to $F_2(N,Q^2)\equiv \int_0^1 \!dx\, x^{N-1} F_2(x,Q^2)$
(and analogously 
for parton distributions). 

Separation of light flavours and antiflavours can be obtained by
including different observables along with the DIS structure
functions~\cite{mrst,cteq} in the set of processes used to determine
the pdfs. Specifically, light antiflavour separation
is obtained comparing  (Tevatron) Drell-Yan production with proton or deuteron
targets, because
\beq
\frac{\sigma^{pd}}{\sigma^{pp}}\Bigg|_{x_1>>x_2}\approx\frac{1}{2}\left(1+\frac{\bar
  d(x_2)}{\bar u(x_2)}\right),
\label{dyas}\eeq 
and light flavour separation from (Tevatron)
 $W^\pm$ production asymmetry data,
which, neglecting strange quark contributions, give
\beq
\frac{\sigma_{W^-}}{\sigma_{W^+}}\approx\frac{d(x_1) \bar u(x_2)}{
u(x_1) \bar d(x_2)}.
\label{was}\eeq 
Inclusion of these data is mandatory in order to reduce the
uncertainty on quark and antiquark distributions significantly below
10\% (see fig.~2), as required for LHC phenomenology.

Strangeness can only be
determined through weak current DIS, i.e., essentially neutrino data
(and some HERA data), which are however scarce and subject to sizable
uncertainties. As a consequence, in current
fits~\cite{mrst,cteq,ale} 
the shape of the
strange distribution is not determined, and 
assumed to be related in a fixed way to that of the light quark sea;
the $s$ and $\bar s$ distributions have been determined only in 
dedicated analyses based on neutrino data~\cite{neufits}.
\begin{figure}[ht]
\vspace{-1.2cm}
\includegraphics[width=.49\linewidth,clip]{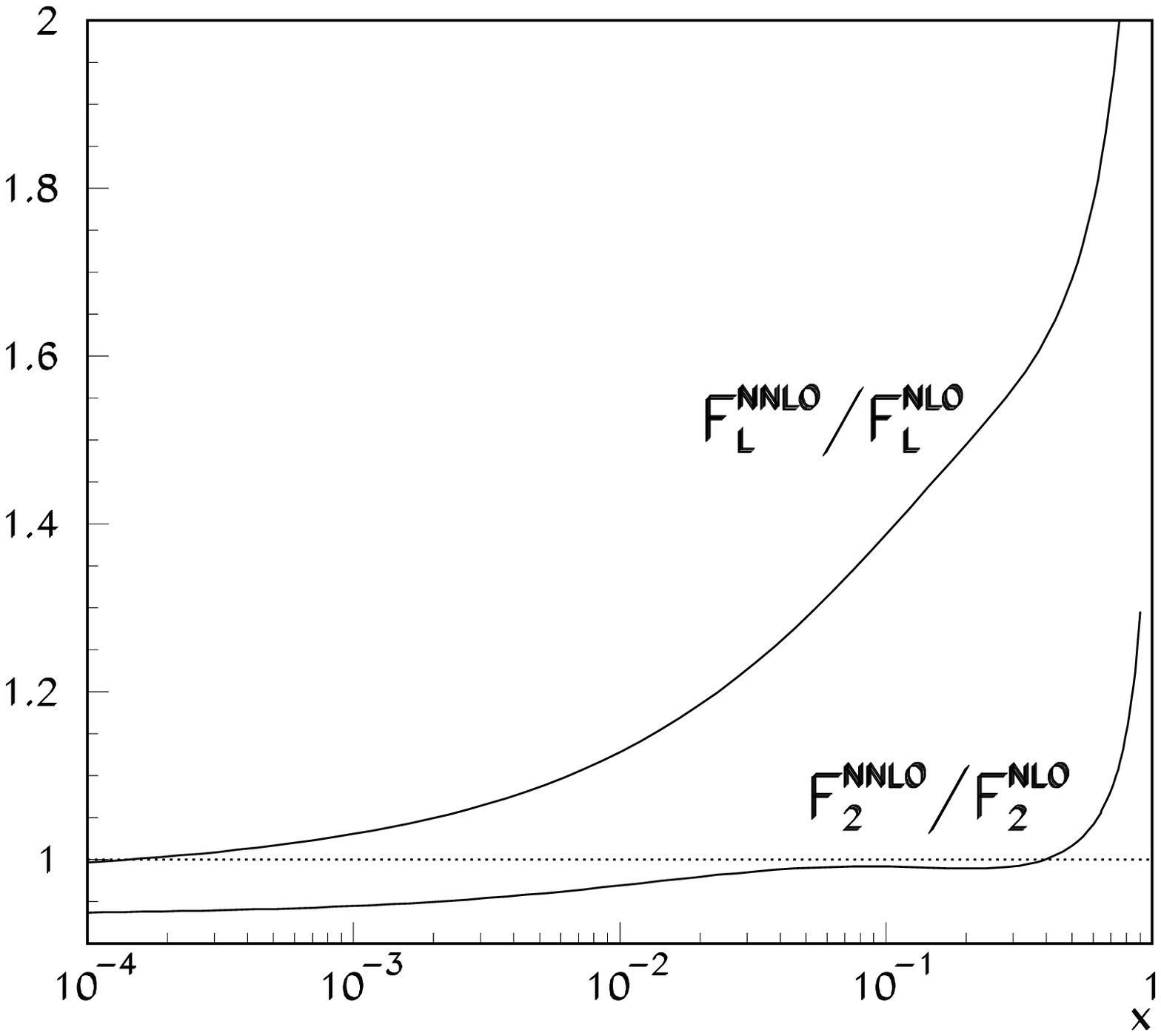}
\includegraphics[width=.49\linewidth,clip]{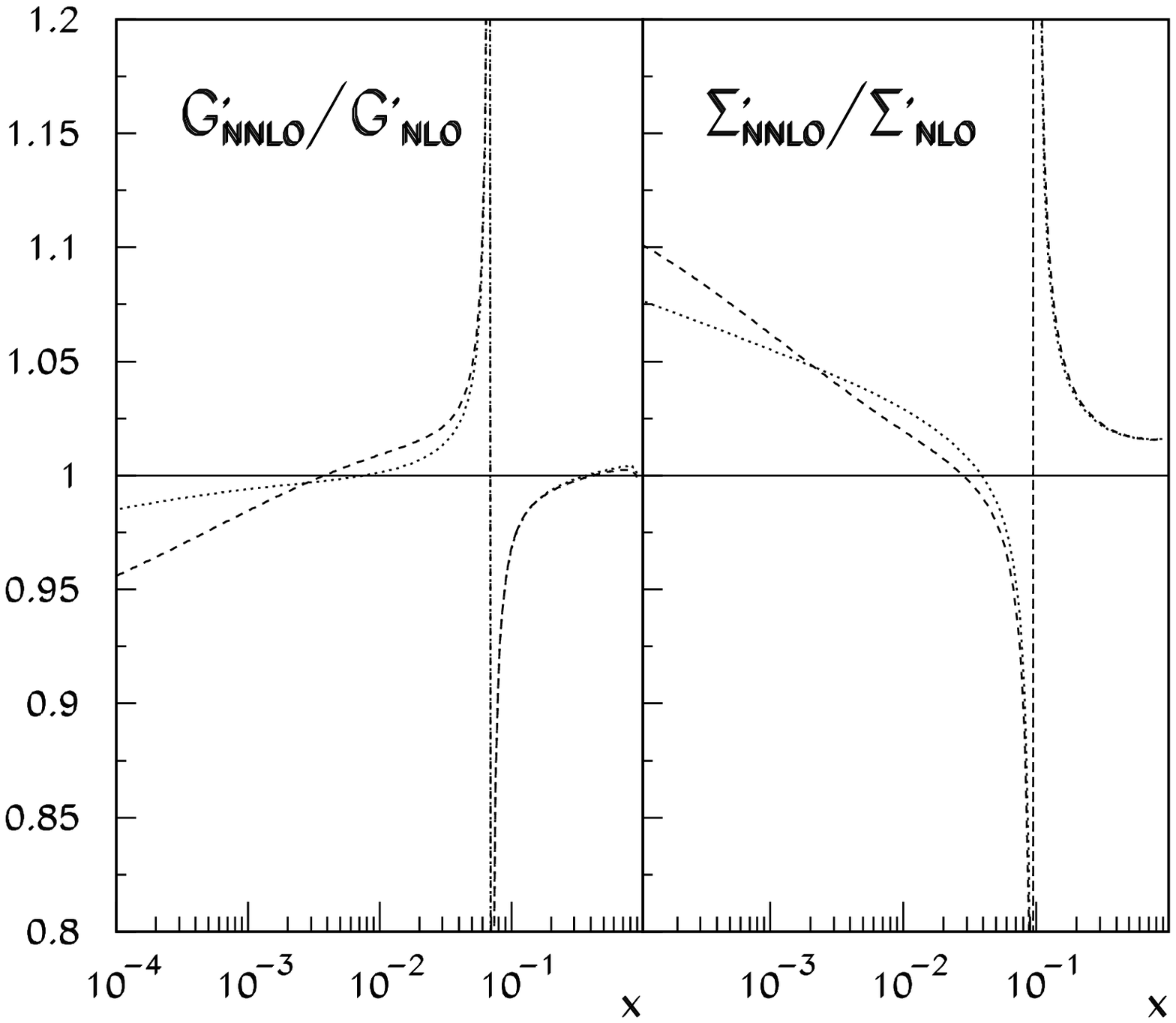}
\vskip-1.6cm
\caption{Impact of the NNLO corrections on (left) the coefficient functions
  and  (right) the evolution of parton distributions. (from Ref.~\cite{aleimp})}
\label{nnloimp}
\vspace{-.5cm}
\end{figure}

Finally, because (see fig.~2)   $\gamma_{qg}\ll\gamma_{qq}$ at large $N$, the
gluon can only be determined accurately from eq.~(\ref{f2ev}) at small $N$,
i.e., inverting the Mellin transform, at small $x$. In some parton
sets~\cite{mrst,cteq} accuracy on the gluon
at large $x$ is improved through the inclusion of 
data from inclusive jet production at the
Tevatron. 

A better handle on the gluon could be obtained if the contributions of
$F_2$ and $F_1$ to the cross section eq.~(\ref{disxsec}) could be
disentangled. This is
possible by varying $y$ at fixed $x$ and
$Q^2$, which requires varying the beam energy: it would be possible
by lowering the beam energy at HERAII. On top of this, high
luminosity and especially weak scattering data from HERAII could
improve somewhat current
flavour separation~\cite{heraII}. It is unclear whether any of these
measurements might be performed before the shtudown of HERA.

A markedly more significant improvement in knowledge of parton
distributions is expected from the coming into operation of
the LHC itself, even though only extremely limited studies of the
impact of LHC on PDFs are available so far~\cite{lhcpdfs}. At longer
term, full information on the flavour decomposition of the nucleon
could come from a neutrino factory~\cite{nufact}. 

\subsection{Theory: perturbative coefficients and evolution}
However abundant the data, pdfs can be extracted from them only
through the use of perturbative parton cross sections which are needed
to relate them to physical obervables and the
anomalous dimensions which are required to evolve them to a common
scale. Hence, the theoretical uncertainty on pdfs is always at least
as large as the size of the unknown perturbative
corrections. Until very recently, only inclusive Drell-Yan and DIS
partonic cross sections  were known at NNLO~\cite{dydisnnlo}. In the
last couple of years, thanks to the development of new
computational techniques, NNLO results have been obtained for 
inclusive Higgs production~\cite{hnnlo} and,
more importantly, for a number of less inclusive observables,
specifically Drell-Yan and W rapidity distributions in hadronic
collisions~\cite{diffnnlo}. 
On the other hand, the full set of NNLO anomalous dimensions or splitting
functions has also been
determined~\cite{splnnlo} after an effort of more than a decade.

The impact of NNLO corrections to DIS coefficients and evolution is
displayed in fig.~\ref{nnloimp}. Clearly, their inclusion is required
if one aims at achieving a determination of parton distributions to
percent accuracy.  Their impact on specific observables (such as
$F_L$) or in particular kinematic regions (such as very small or very
large $x$) can be even more dramatic.
The large size of NNLO corrections at small and large $x$
signals the need for all--order resummation of the perturbative
expansion in these regions. Indeed, higher order perturbative
correcyions are known to be 
enhanced by logs of $x$ and $(1-x)$ through terms of
the form $\alpha_s\ln\frac{1}{x}$ and $\alpha\ln^2(1-x)$.
When $x$ is small or large enough the log enhancement can offset the
suppression due to the strong coupling $\alpha_s$. 

\begin{figure}[ht]
\vspace{-1.cm}
\includegraphics[width=.57\linewidth,clip]{nnnlonfz.ps}
\hskip.04cm
\includegraphics[width=.42\linewidth,clip]{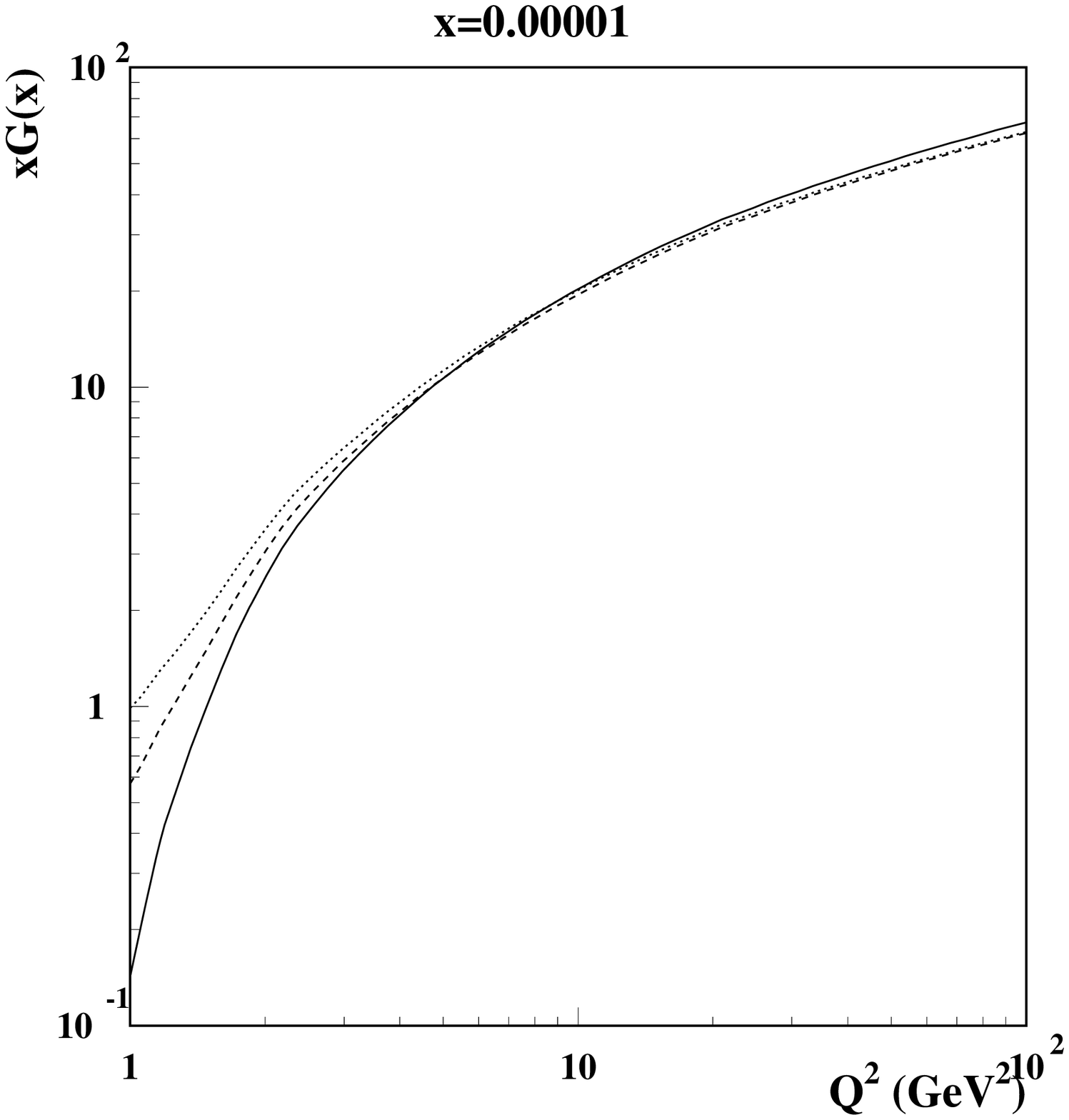}
\vskip-.8cm
\caption{Left: the splitting function ($P_{gg}$ with $n_f=0$) at LO
  (solid, black), NLO (dotdashed, green), NNLO (dashed, red), NNLO
  leading singularities (lower dotted, red), NNNLO leading
  singularities (upper dotted, blue). The two solid (blue) curves with
  a dip are the small $x$ resummations of Ref.~\cite{abf} and Ref.~\cite{ciaf}. 
Right: best-fit 
gluon at NLO
  (bottom, solid), NNLO (middle, dashed), and NNLO evolution of the
  NLO best-fit (top, dotted, from Ref.~\cite{aleprel}).}
\label{smallx}
\vspace{-.5cm}
\end{figure}
The impact of large $x$ corrections beyond NNLLO on the extraction of
pdfs from DIS has been estimated~\cite{mrstun} 
to be negligible at least if one
imposes a cut on
the center-of-mass energy of the $\gamma^* p$ 
process $W^2\equiv\frac{Q^2(1-x)}{x}\gsim 10$~GeV$^2$.
On the other hand, once less inclusive observables (e.g. the
differential Higgs production cross section) are considered and
experimental cuts are taken into account the impact of large $x$
corrections can become sizable~\cite{grazhiggs}. Because large $x$
resummations are known exactly and their inclusion poses no problem of
principle it would be advisable to include them in future parton sets.

The impact of small $x$ corrections is dramatically highlighted by
the recent NNLO splitting function determination. Indeed 
(fig.~\ref{smallx}) the perturbative expansion of the splitting function
is unstable at  small $x$, but on the other hand the logarithmically
enhanced small $x$ terms (leading singularities) are not a good
approximation to the full result. This means that small $x$
contributions have to be resummed, but this resummation must also be
combined with the available fixed-order results in order to lead to a
stable answer. The required formalism has been developed over the last
decade by various groups and is now converging to an answer, but the
relevant phenomenology has not been developed yet. Current
results~\cite{abf,ciaf} show (see fig.~\ref{smallx}) 
that the resummation stabilizes the NLO
results, so that at small $x$ the fully resummed result is actually
closer  to the low order (LO and NLO) results. 
The impact of
NNLO corrections on the extracted gluon can be larger than 100\% at
small $x$ and small scale (see fig.~\ref{smallx}).
Hence, the resummation
is mandatory if one wishes to work to NNLO accuracy. 
\section{Partons and errors }
Precision phenomenology needs not only a knowledge of parton
distributions, but also of the error with which they are
determined. 
 The
problem here is that parton distributions are functions, hence the
error on them is really a probability measure in an
infinite-dimensional space. Therefore, it cannot be
determined from a finite set of data without extra theoretical
assumptions. These assumptions in current parton fits take the form of
a functional form: pdfs are assumed to have, at a
reference scale,  a given
functional form, parametrized by a finite set of parameters. They are
then evolved to the scale of the data and used to compute physical
cross--sections which are then fitted to the data, thereby determining
the parameters. If the full information of the covariance matrix of
the data is retained and propagated to the parameters it is then
pssible to determine errors on pdfs.
Within the last few years, three parton sets with errors
have been obtained in this way~\cite{mrst,cteq,ale}.

\begin{figure}[ht]
\vspace{-1.cm}
\includegraphics[width=.49\linewidth,clip]{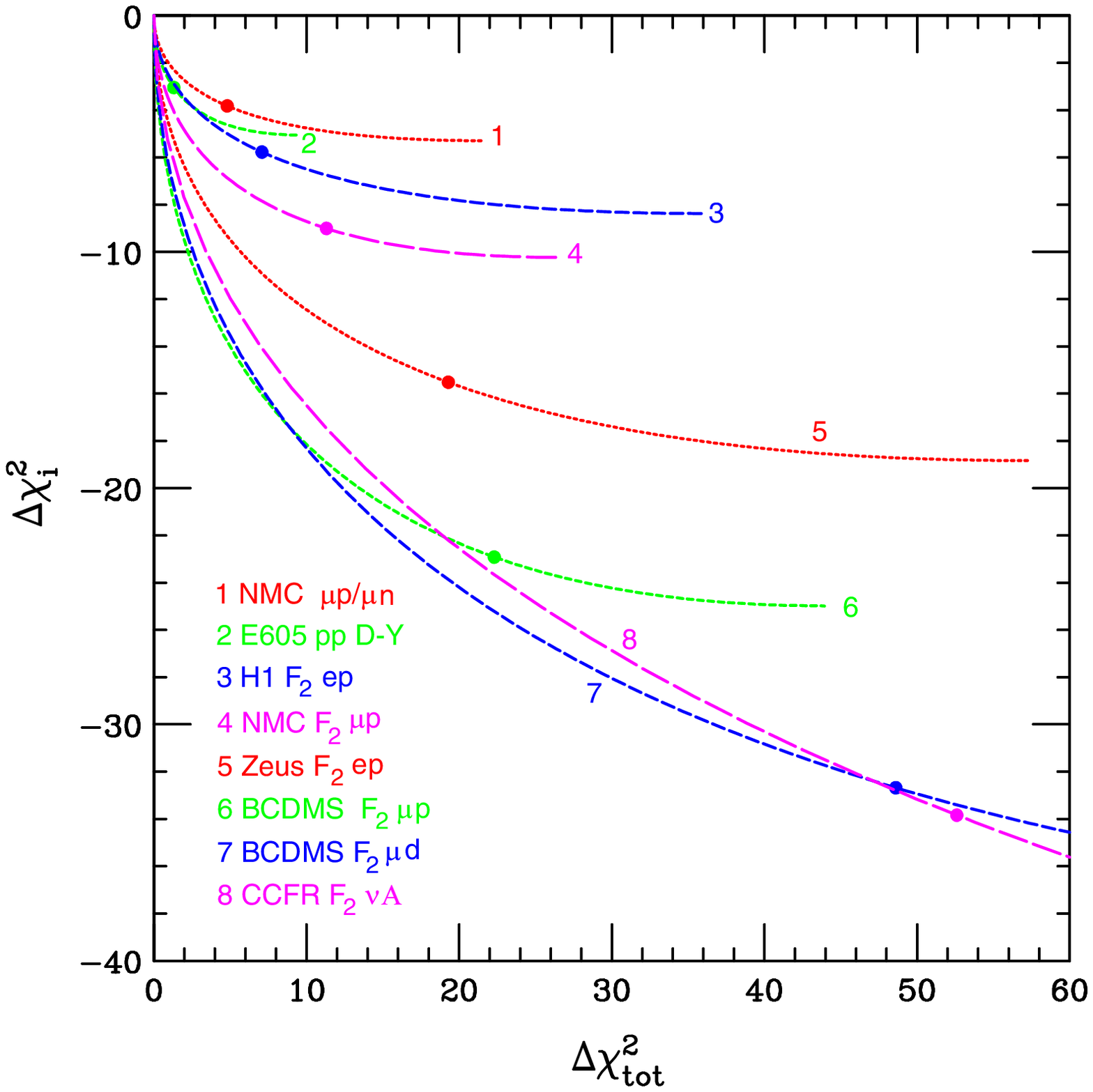}
\includegraphics[width=.49\linewidth,clip]{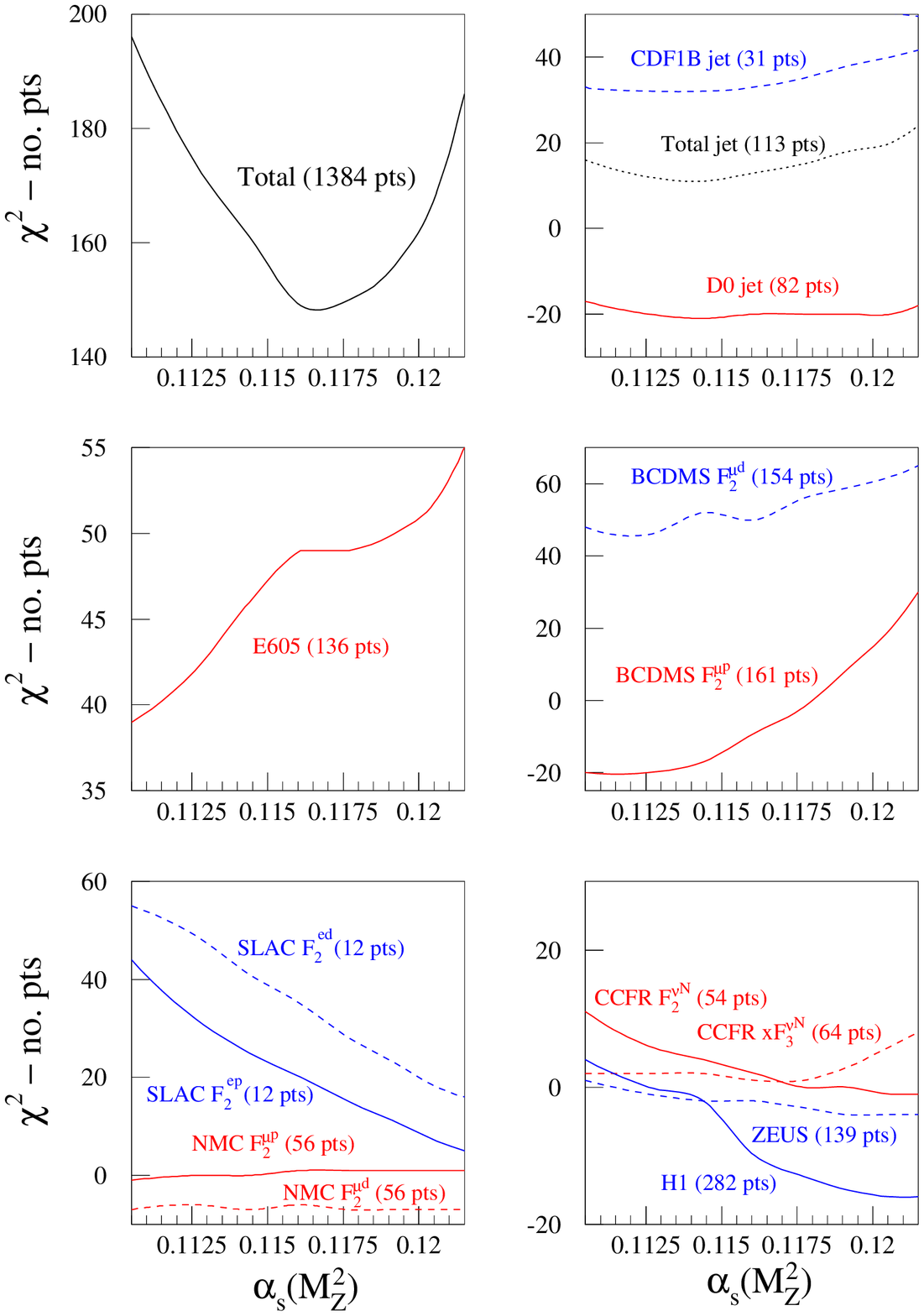}
\vskip-.8cm
\caption{Left: minimum $\chi^2$ for individual experiments as the
  total $\chi^2$ is allowed to increase (from Ref.~\cite{colpump} ). 
Right: contribution to the $\chi^2$ of the global MRST fit
  from individual experiments (from Ref.~\cite{mrst})
}
\label{fitprob}
\vspace{-.9cm}
\end{figure}
\subsection{The problems}
The result of current global parton fits look nominally quite good,
with uncertainties of order of a few percent on quark and antiquark 
distributions and at most 10\% on the gluon (see fig.~2). However,
closer inspection reveals a number of problems. 
Indeed, consider  the variation of the $\chi^2$
of the fit to each dataset as the $\chi^2$ of the global fit is allowed
to vary~\cite{colpump}. This study reveals that the
fit to the CCFR neutrino DIS data or the BCDMS muon-deuterium DIS data
can be improved very considerably at the expense of deteriorating the
global fit (fig.~\ref{fitprob}). This indicates that the global fit
with the given functional form is far from the best fit to these data.
Also, one may study the contribution of individual datasets to the
$\chi^2$ of the global fit as one of the fit parameters is varied.
This  (fig.~\ref{fitprob}) shows that the 
minimum of the global fit  does not correspond to minima of individual
experiments.

While to some extent these could just be statistical
fluctuations, they may signal more
serious problems. On the experimental side, they may signal that
some datasets are not fully
consistent with the others. 
On the theoretical side, they may signal that 
assumptions on the functional form are not flexible enough to
accommodate the data.

\begin{figure}[ht]
\vspace{-1.cm}
\includegraphics[width=.49\linewidth,clip]{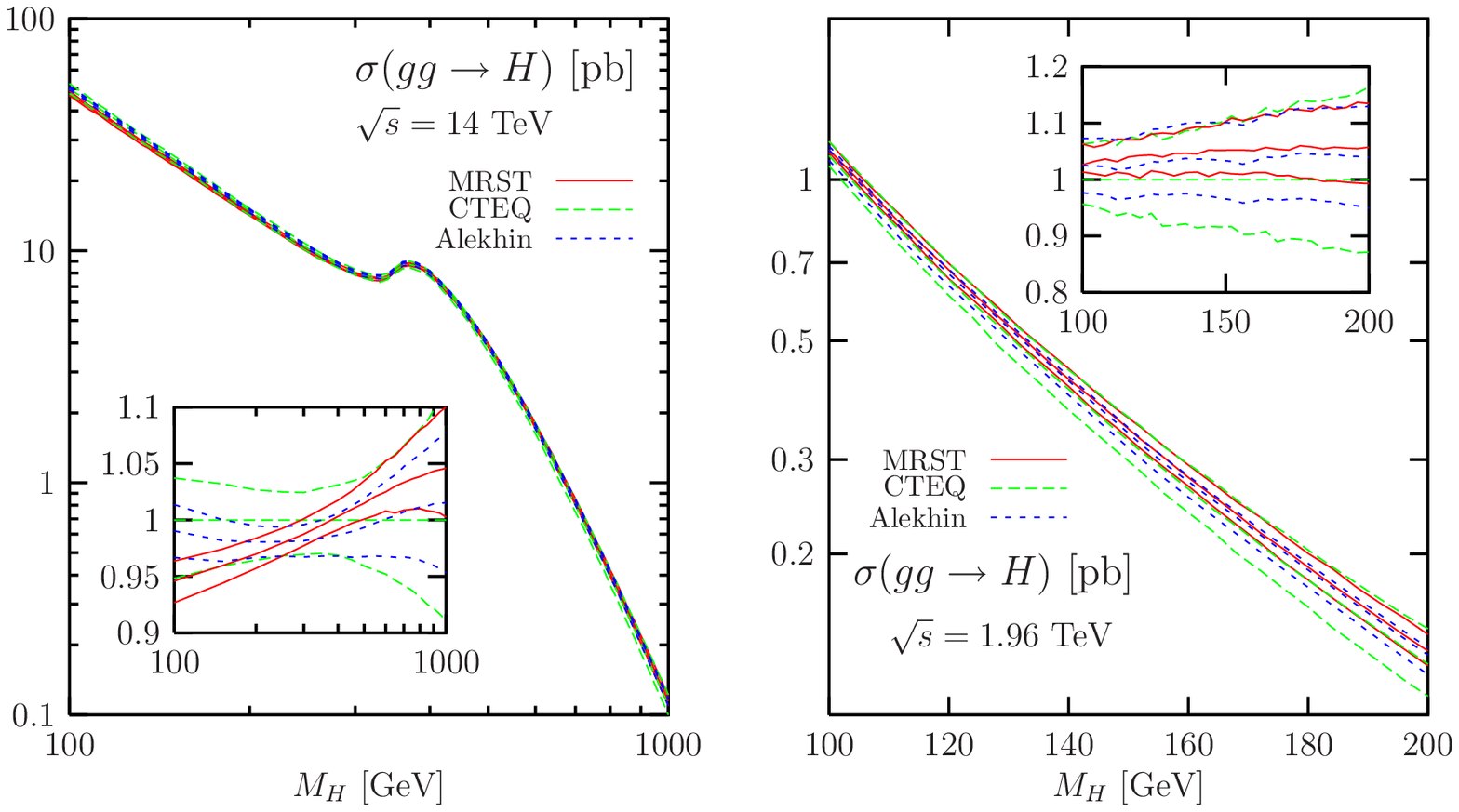}
\includegraphics[width=.49\linewidth,clip]{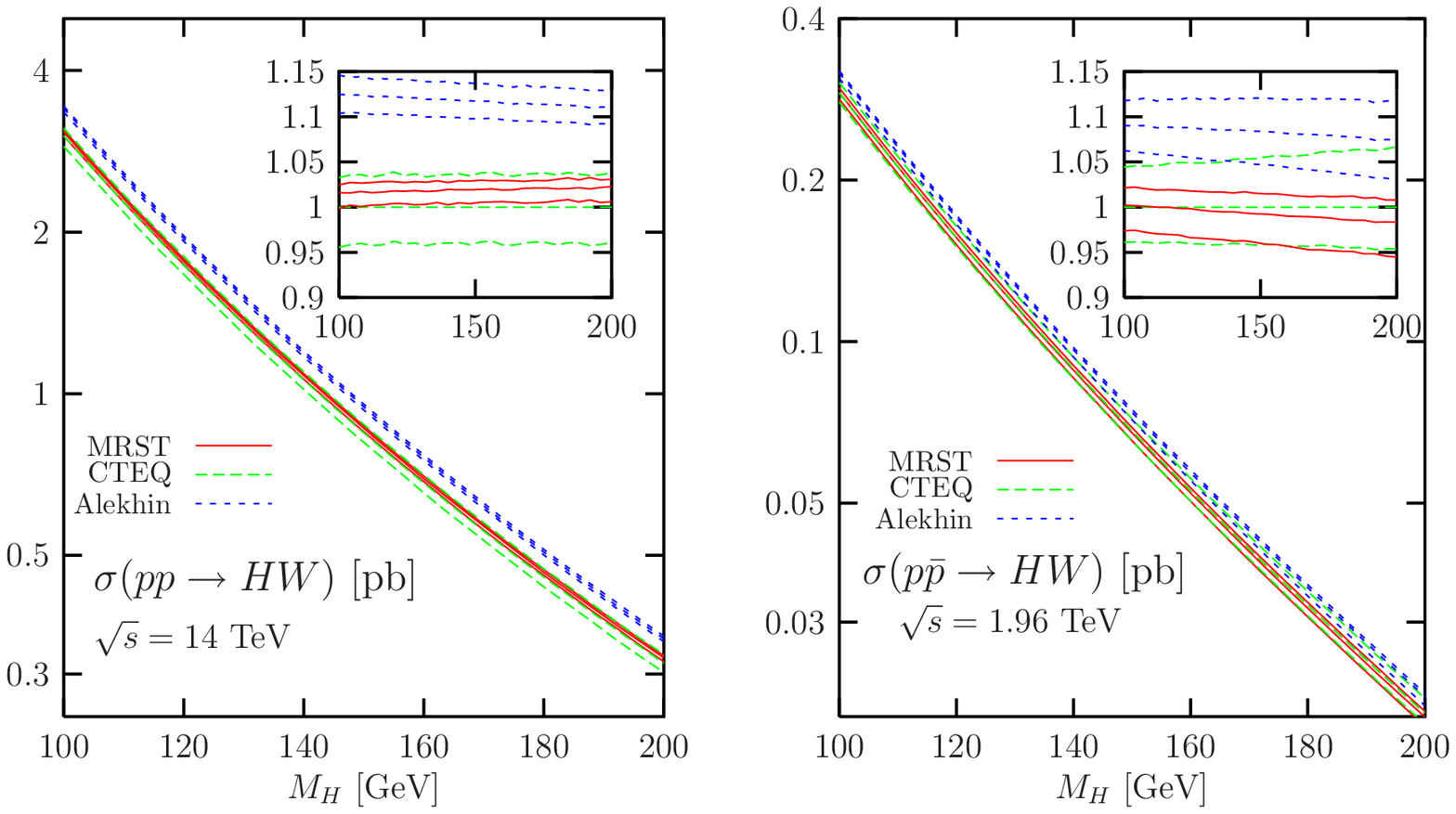}
\vskip-.8cm
\caption{ Higgs production cross section at the LHC computed using
  various parton sets (from Ref.~\cite{df})
}
\label{higgsxs}
\vspace{-.5cm}
\end{figure}
Whatever the precise cause, it appears that current determinations of
errors on pdfs have not yet settled to a  satisfactory agreement. This
is dramatically apparent if one looks at the prediction for even the
simplest inclusive LHC observable, such as the total Higgs production
cross section (fig.~\ref{higgsxs}): the results obtained using recent
parton sets disagree within the respective error bands, expecially when
the quark flavour separation comes into play, such as in $HW$ production.

\begin{figure}[ht]
\vspace{-1.cm}
\includegraphics[width=.49\linewidth,clip]{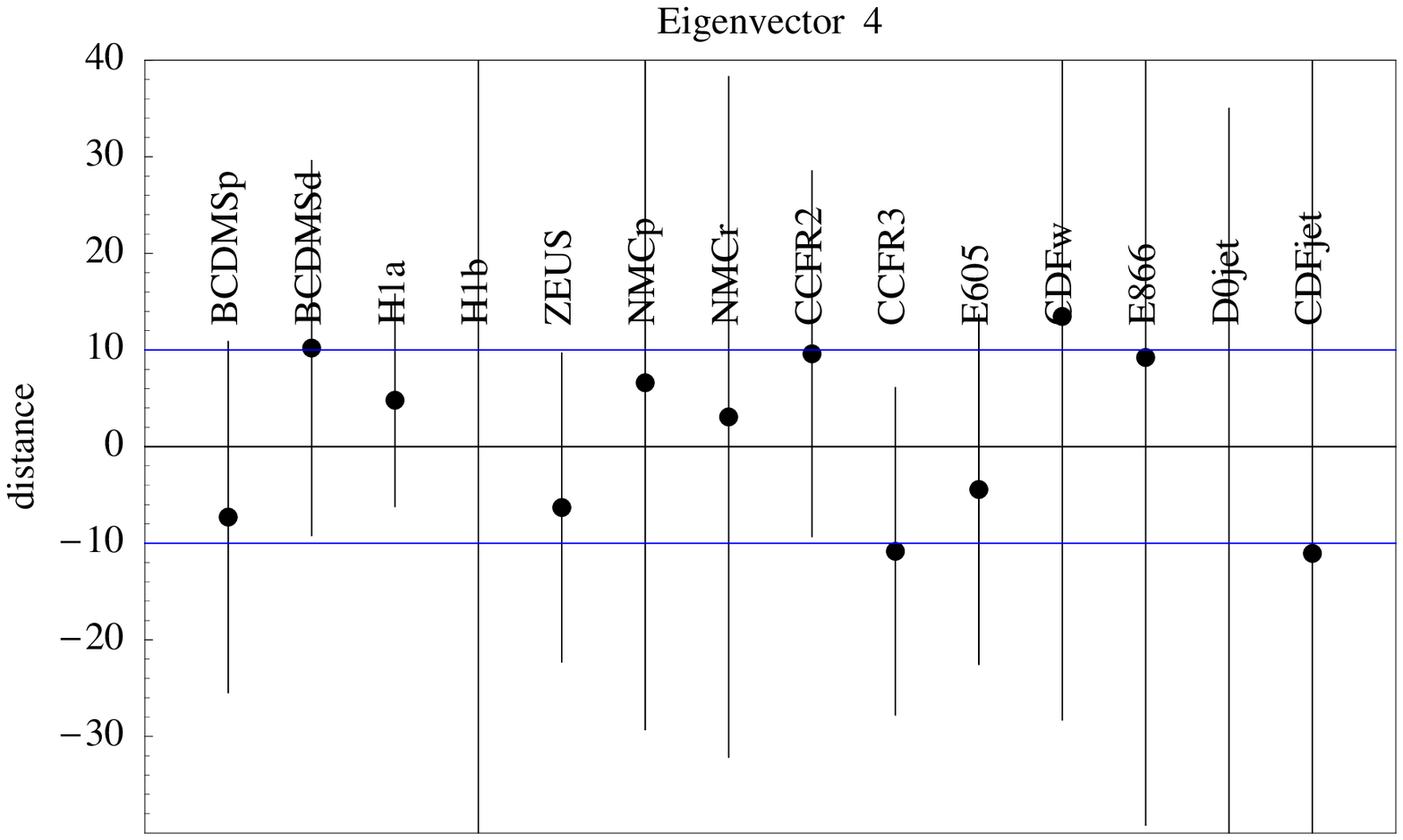}
\includegraphics[width=.49\linewidth,clip]{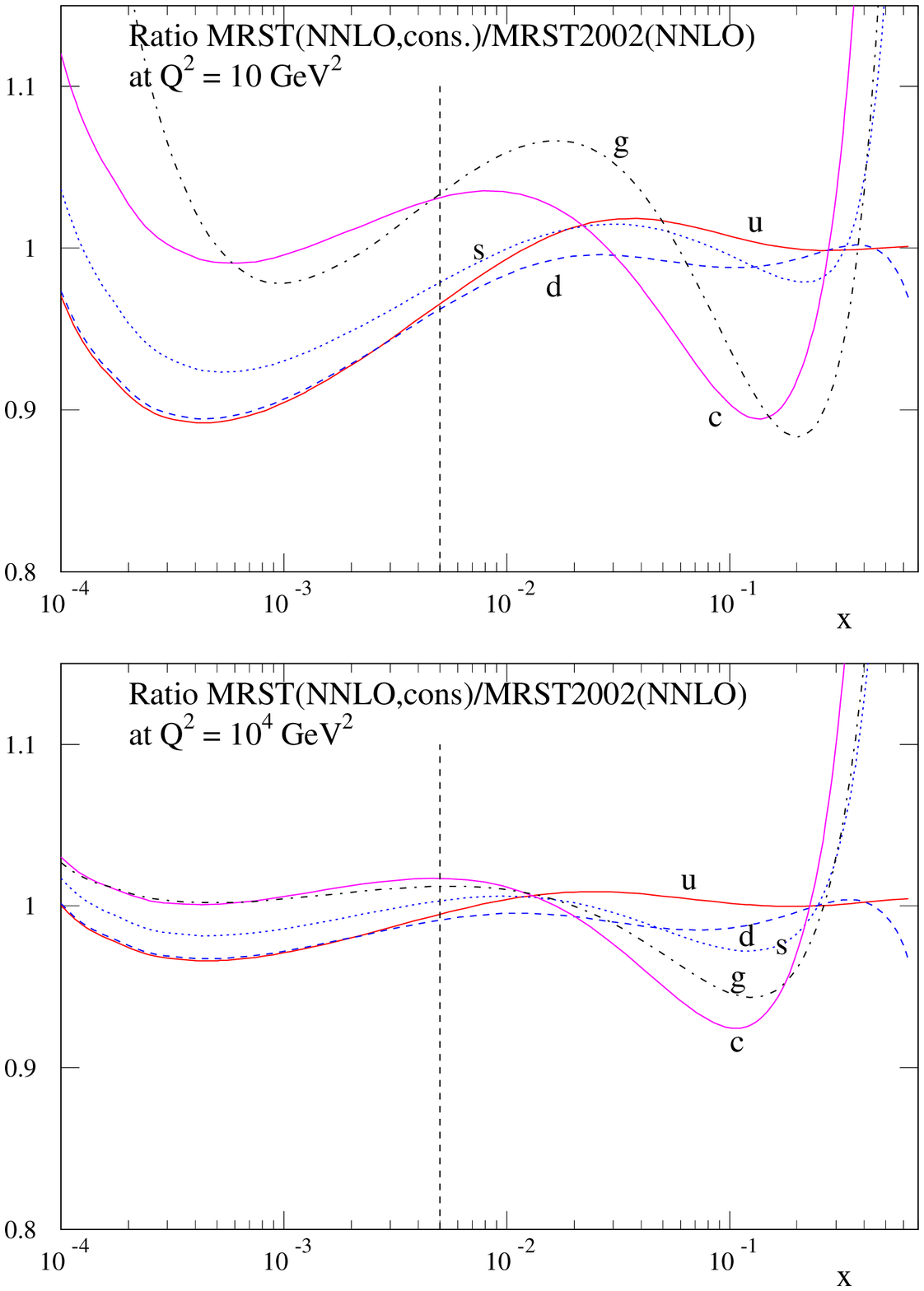}
\vskip-.8cm
\caption{Left: 90\% probability ranges for each
  experiment as a function of the distance from the minimum along a
  typical eigenvector of the covariance matrix. The blue band
  corresponds to $\Delta \chi^2=100$ (from Ref.~\cite{cteq}). Right: ratio of
  MRST `conservative' partons to the reference 
  MRST set (from Ref.~\cite{mrst} ). }
\label{cons}
\vspace{-.5cm}
\end{figure}
\subsection{Conservative solutions}
The problems of data incompatibility and possible parametrization bias
have been tackled in various ways in current parton sets. A first
option is to replace the standard one-sigma contours with parameter
ranges obtained studying the compatibility of the fit with various
experiments. In ref.~\cite{cteq} this has been done by studying the
spread of 90\% confidence intervals for various experiments, as one
moves away from the minimum of the $\chi^2$ along eigenvectors of the hessian
matrix, and taking the envelope of the resulting ranges. In practice,
this suggests that $\Delta \chi^2=100$ for the global fit
leads to a reasonable estimate of the one-sigma contours for
pdfs (see fig.~\ref{cons}).
In ref.~\cite{mrst}  $\Delta \chi^2=50$ is adopted instead,
and seen to lead to results which are not so different for the pdf
error bands. 

However, the need to chose ad-hoc a large value of $\Delta \chi^2$
is somewhat disturbing. An alternative suggestion has been made 
in Ref.~\cite{mrst}, where
it is observed that most of the trouble
seems to come from specific kinematic regions where theoretical
uncertainties become large: the low $Q^2$ region where the
perturbative expansion converes slowly, and the large and small $x$
regions where resummation is necessary. It is then shown that by
imposing more restrictive cuts in $Q^2$, $x$ and $W^2$ (see sect.~2.2
above) a much more palatable value $\Delta \chi^2=5$ can be taken to
determine the error on pdfs. The `conservative'
partons obtained in this way can differ by more than 10\% from
standard ones (see fig.~\ref{cons}). 
The problem is that clearly  there is information loss
in the procss, and predictions become unreliable when regions are
probed which have been excluded from the fit due to the cuts (such
as the very small $x$ region).

Still, one would like to be able to rely on purely statistical
arguments to construct one sigma contours.
To this purpose, in ref.~\cite{ale} it has been observed that
many problems seem to
come from the need of combining different data sets. Indeed,
in ref.~\cite{ale} it has ben demonstrated
 that if only DIS data are included in the global
fit, and the full covariance matrices of experiments are taken into
accout, it is possible to achieve a statistically stable fit where
one-sigma error bands are given by $\Delta \chi^2=1$. The problem is
that DIS data alone are insufficient to determine for instance the
quark flavour decomposition. However,  
preliminary results~\cite{aleprel} suggest that
 DIS data can be combined e.g. with
Drell-Yan data, provided only that the  $\chi^2$ from different
dataset are brought to a common normalization. Satisfactory 
errors on all pdfs are then 
obtained (see fig.~2). 

The fact that different prescriptions seem to be able to solve at
leats in part the problems of global fitting suggests that the origin
of these problems is not yet fully understood.

\begin{figure}[ht]
\vspace{-1.cm}
\includegraphics[width=.49\linewidth,clip]{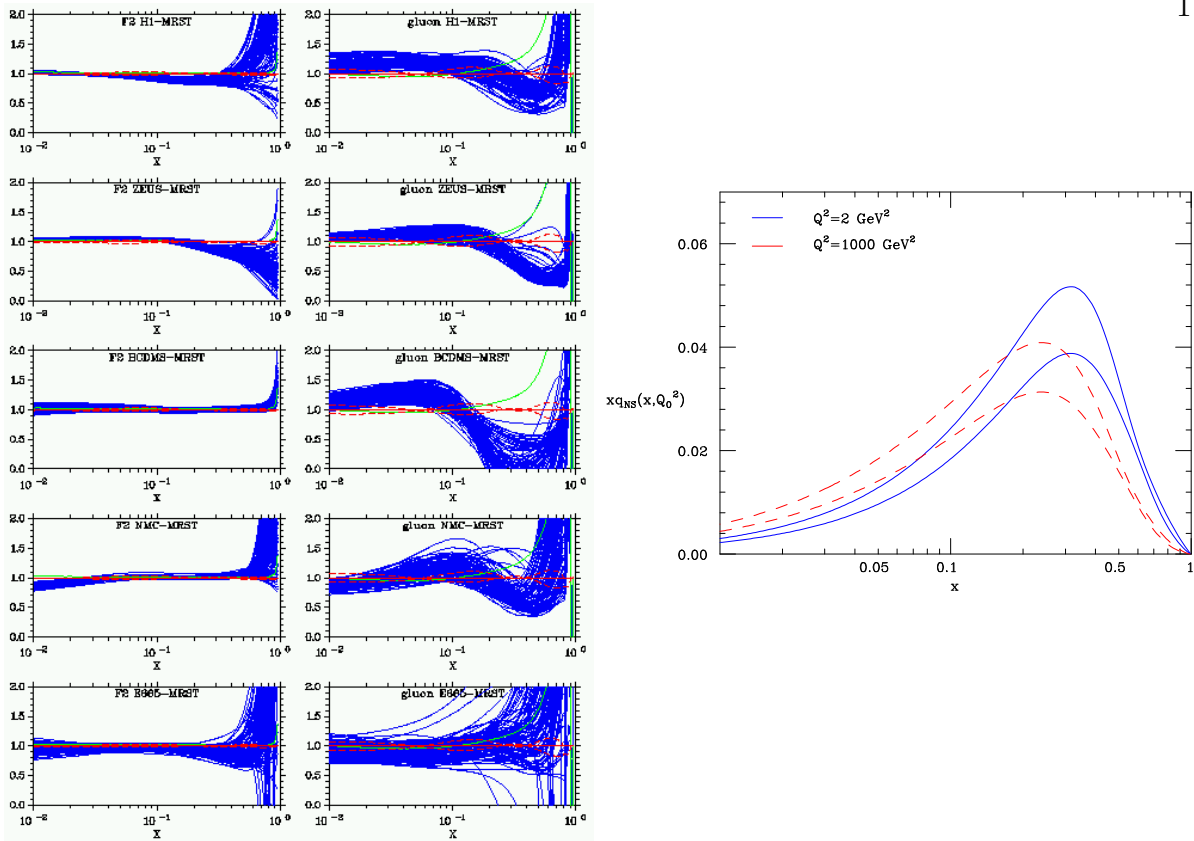}
\includegraphics[width=.49\linewidth,clip]{qns.ps}
\vskip-.8cm
\caption{Left: Fermi partons (from Ref.~\cite{gkk}). Right: nonsinglet neural
  partons (preliminary,~\cite{nnqp}).}
\label{gkknn}
\vspace{-.5cm}
\end{figure}
\bigskip
\subsection{New ideas}
The difficulties encountered in current  parton fits
suggest that perhaps the conventional approach is now reaching its
limitations, and has led to the suggestion of alternative approaches. A
first proposal~\cite{gkk} is to use Bayesian inference to update a
prior representation of the probability density which is
generated as a Monte Carlo
sample based e.g. on an available parton parametrization. 
The final result should be largely independent
of the choice of prior if the data are sufficiently abundant. The
main difficulties with this approach are related to the need 
to keep the computational complexity under control, in particular in 
the choice of priors and in the
handling of
flat directions, i.e. Monte Carlo replicas which lead to similar
values of the $\chi^2$. A preliminary set of partons ('Fermi' partons)
has been contructed within this approach~\cite{gkk} (see
fig.~\ref{gkknn}). The results suggest that  indeed a treatment of non-gaussian
probability densities may be required if one wishes to combine
experimental information from different sets. However, no satisfactory
global fit within this approach has been obtained yet: in particular,
the preliminary results do not lead to a satisfactory value of the
strong coupling.

Another approach has been suggested in ref.~\cite{neuroquarks}, based
on the idea of using neural networks as universal unbiased
interpolants. In this approach, the data are used to generate a Monte
Carlo sample which represents the probability measure in the space of
functions at the points at which data exist. Neural networks are then
used to interpolate between these points: the ensuing Monte Carlo set
of neural networks is then the sought-for probability in the space of
functions. This approach has been used in ref.~\cite{neuroquarks} to
parametrize all available $F_2$ data, but without extracting the
contribution of individual pdfs to the structure function. Preliminary
results on a pdf extraction based on this method have been presented
in ref.~\cite{nnqp} (see
fig.~\ref{gkknn}). They suggest that fixed functional forms may be too
rigid in estimating errors expecially at the edges of the data
region. The feasibility of a full parton set based on this approach,
which is also computationally quite intensive,
is however still to be demonstrated.

These approaches have in common the feature of trying to use the
available experimental information in a way which is free of
theoretical assumptions. 
\section{Conclusions}
Perturbative QCD phenomenology has become the object of precision
quantitative studies during the last decade and it is now on a similar
footing as precision electroweak phenomenology. However, unlike in
the electroweak case, the
impossibility to compute the structure of the nucleon from first
principles poses taxing problems of data analysis. A satisfactory
agreement between different determination of parton distributions has
not been reached yet especially at the level of error determination,
and may require the development of entirely new techniques.
\bigskip

{\bf Acknowledgements}: I thank B.~Pire and M.~Guidal for
inviting me to this stimulating meeting,  R.~Ball and G.~Ridolfi for
several discussions, and S.~Alekhin for communicating the unpublished
plots shown in figs. 2-4.

\end{document}